\documentclass[11pt,twoside]{article}
\usepackage{vdbbook}
\usepackage{natbib}
\usepackage{times}  
\usepackage{listings}
\usepackage{graphics}
\usepackage{subfigure}
\usepackage{graphicx}
\usepackage{amssymb}
\usepackage{hyperref}
\usepackage{aas_macros}

\resetcounters

\begin{document}

\allowtitlefootnote

\title{Simulations and modelling of the interstellar medium in galaxies}
\author{Claudia del P. Lagos$^{1}$, Cedric G. Lacey$^2$, Carlton M. Baugh$^2$
\affil{$^1$European Southern Observatory, Karl-Schwarzschild-Strasse 2, 85748, Garching, Germany.\\
$^2$Institute for Computational Cosmology, Department of Physics,
University of Durham, South Road, Durham, DH1 3LE, UK.}}

\begin{abstract}
The latest observations of molecular gas and the atomic hydrogen content of local 
and high-redshift galaxies, coupled with how these correlate with star formation activity, 
have revolutionized our ideas about how to model star formation in a galactic context. 
A successful theory of galaxy formation has to explain some key facts: 
(i) high-redshift galaxies have higher molecular gas fractions and star formation 
rates than local galaxies, (ii) scaling relations show that the atomic-to-stellar 
mass ratio decreases with stellar mass in the local Universe, and (iii) 
the global abundance of atomic hydrogen evolves very weakly with time. 
We review how modern cosmological simulations of galaxy formation attempt to put 
these pieces together and highlight how approaches simultaneously solving 
dark matter and gas physics, and approaches first solving 
the dark matter $N$-body problem and then dealing with gas physics using semi-analytic models, 
differ and complement each other. We review the observable predictions, what we think we have learned 
so far and what still needs to be done in the simulations to allow robust testing 
by the new observations expected from telescopes such as ALMA, PdBI, LMT, JVLA, ASKAP, MeerKAT, SKA.
\end{abstract}

\section{Introduction}
\label{Sec:Intro}

\begin{figure}[t]
\centering
\includegraphics[width=1.0\textwidth]{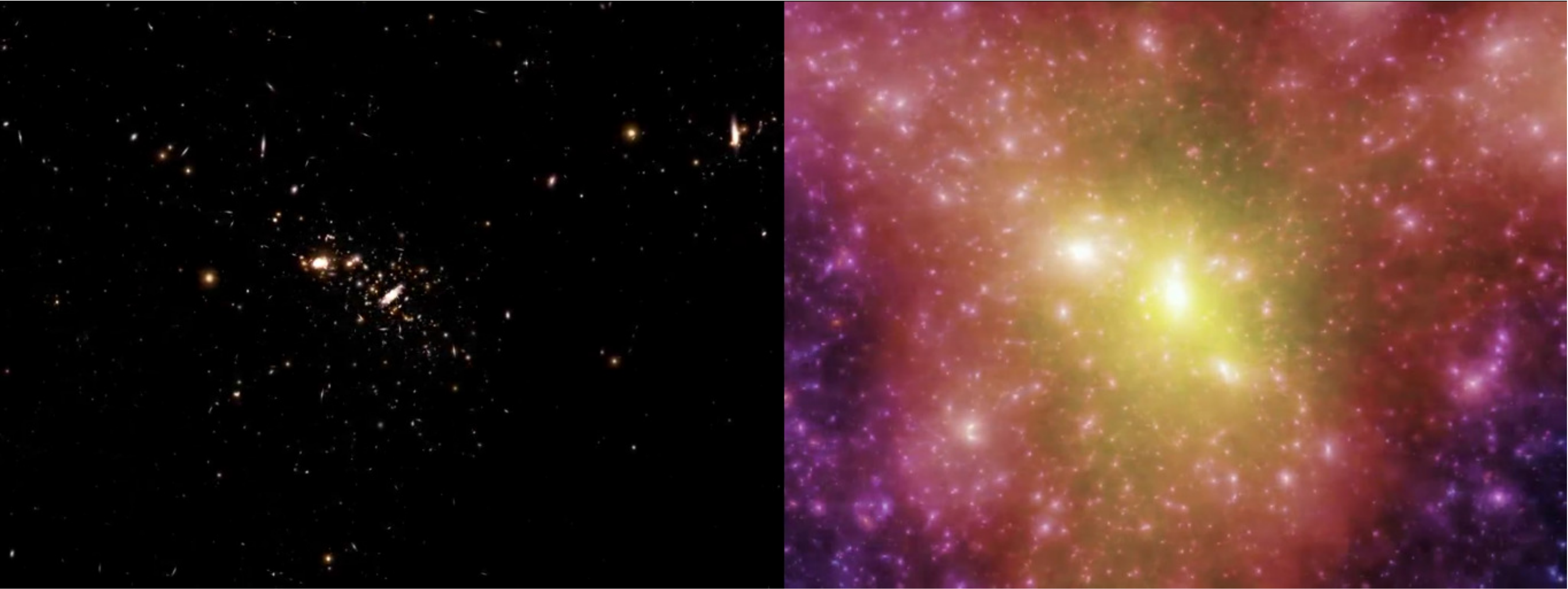}
\caption{Snapshots of a Millennium fly through visualization 
made by Mark Swinbank, Daniel Farrow and Joel Snape
from the Millennium simulation \citep{Springel05} combined with the semi-analytic model {\tt GALFORM}. {\it Left panel:} Large scale view of galaxies in and around a galaxy cluster. Colours and positions 
of galaxies correspond to those predicted by {\tt GALFORM}. 
{\it Right panel:} Same volume as in the left panel but showing the position of the 
DM, where yellow represents higher densities of DM and purple lower densities. The DM visualization is used with the kind permission of 
Volker Springel.}
\label{fig:DMsnap}
\end{figure}

A fundamental challenge in astrophysics is to understand the connection between a very homogeneous early 
Universe, as probed by the Cosmic Microwave Background (CMB) radiation and its temperature fluctuations 
(e.g. \citealt{Komatsu11}), and the density contrasts observed today in 
on large scales, such as the galaxy clustering (\citealt{Tegmark04}), and the small scales, such as stars. 
The wide range of scales associated with to this problem, from a 
few astronomical units to Gpc, is also connected to several orders of magnitude 
in temperature and density. The computational power currently available makes it impossible to treat all the physics relevant 
on the different 
scales self-consistently, making it necessary to treat the problem in pieces. This is the case of galaxy formation, 
in which a variety of physical processes taking place in different scales can have a large impact on the evolution 
of galaxies, such as stellar feedback on small scales to gas infall from filaments to the 
centre of dark matter (DM) halos, where galaxies reside, on large scales. Fig.~\ref{fig:DMsnap} shows a visualisation of
 a DM only simulation and how galaxies populate DM halos as predicted by 
the semi-analytic model of galaxy formation {\tt GALFORM} \citep{Bower06}, 
emphasizing the role the cosmological paradigm plays in the theory of galaxy formation.

To study galaxy formation, it is necessary to simulate the evolution of baryons in the context of the cosmological growth
of the DM structures, as this sets the gas inflow and merger histories of galaxies. 
There are two widely used approaches to study galaxy formation: hydrodynamical simulations or ``parallel'' approaches,
which simultaneously follow the evolution of the
DM and gas physics, and ``serial'' approaches or semi-analytic modelling
of galaxy formation, which first solve the DM $N$-body problem and then deal with the gas physics.
In the case of the hydrodynamical simulations, the main advantage is the sophistication of
the simulations which help to avoid many (but not all) prior assumptions in the gas dynamics (e.g. \citealt{Springel03}).
The main disadvantages are that the number of galaxies that can be simulated is still very limited and that,
despite the sophistication of the
technique, many of the physical processes regulating galaxy formation occur on a scale well
below the resolution of these simulations, and are viewed as {\it sub-grid physics} in this context, and are hence
 treated in a phenomenological way. In the case of semi-analytic models,
a hybrid set of discrete and numerical
physical models are implemented to treat galaxy formation and evolution. Semi-analytic models
are able to simulate large cosmological volumes containing
millions of galaxies over cosmic epochs making multiwavelength predictions
(\citealt{Baugh06}). The main drawback
of semi-analytic models lies in the large set of parameters used to model some of the most
uncertain physics, which is also a problem affecting the sub-grid physics part of
hydrodynamical simulations. The improvement of observational techniques and quality of the available data 
along with more accurate and sophisticated theoretical models of relevant physical processes, such as star formation (SF), 
help to remove some of the uncertainties in cosmological models and simulations of galaxy formation.

\begin{figure}[t]
\centering
\includegraphics[width=0.7\textwidth]{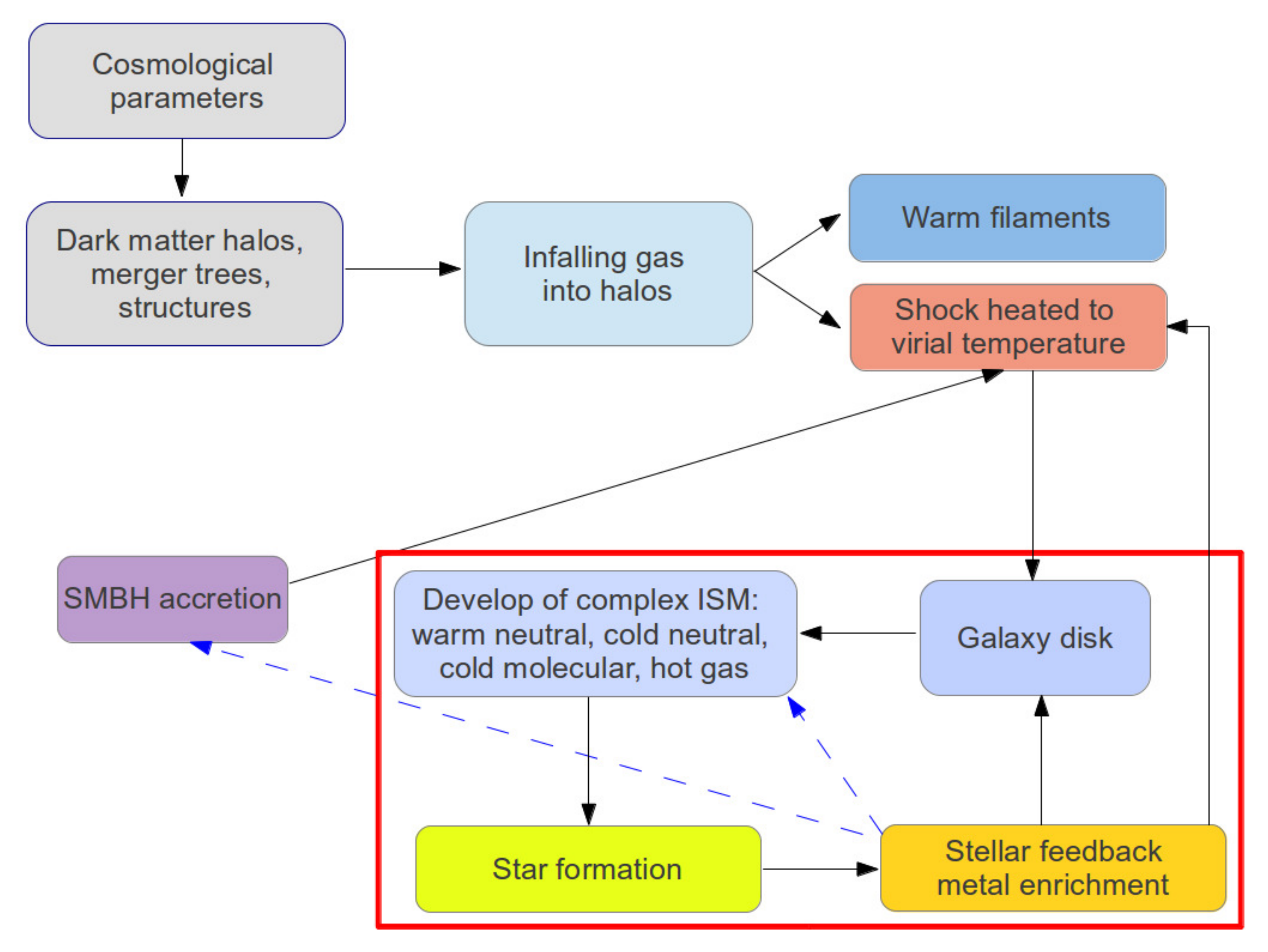} 
\caption{Schematic view of the physical processes involved in the formation of galaxy disks and the ISM.
 Galaxies are assumed
to form embedded in DM halos and therefore the first step is to set the cosmological paradigm. This paradigm defines when and where structures 
form in the universe. Baryonic matter infalls into halos, which have a potential well largely dominated by DM. The gas cooling from shocked heated gas in the halo 
or direct infall of gas to the centre through filaments give rise to galaxies or feeds the existing galactic disk. The chart emphasizes 
the impact the modelling of the ISM and SF can have on the various 
components of a galaxy and even outside the galaxy, such as the intergalactic medium.} 
\label{fig:Chart_ISM}
\end{figure}

The modelling of the ISM in galaxies is of great importance in the theory of galaxy formation and 
even in cosmology; SF takes place in high density gas, which is 
primarily molecular. Star-forming regions are the places where the most massive stars reside and therefore where 
most of the energy and radiation feedback from stars are released. Large scale feedback in the ISM and outflows 
driven by supernovae (SNe) depend on the nature of the star-forming 
regions (i.e. density of the medium, star formation rate, etc.; \citealt{Efstathiou00}). 
 Outflows induced by SNe feedback might have a large impact on the chemical enrichment of the intergalactic medium 
(e.g. \citealt{Oppenheimer06}; see \citealt{Putman12} for a recent review) 
and on driving turbulence in the ISM \citep{Dobbs11}. The turbulence in the ISM can also 
have a direct impact on the infall of gas towards the centre of galaxies where supermassive black holes reside and consequently on 
the accretion and mechanical power of such black holes. A schematic view of the interplay 
of all the processes above is shown in Fig.~\ref{fig:Chart_ISM}, where the red square encloses all the processes directly depending on the 
modelling of the ISM. The schematic stresses the effect the modelling of the ISM can have on all the properties of galaxies and even on the 
gas outside galaxies. 

In this review we summarise key observational results that have pushed the development of more 
realistic ISM and SF modelling in galaxy formation in $\S 2$, 
some important results from multi-phase hydrodynamical simulations in $\S 3$ and semi-analytic 
models of galaxy formation and the main lessons that can be taken 
from them in $\S 4$. We present conclusions in $\S 5$.

\section{Characterisation of the star formation law and the ISM in galaxies}

Large improvements in the resolution and quality
of the imaging and spectroscopy of nearby galaxies have allowed a better understanding  
of one of the key physical processes in galaxies: star formation. This new set of observations has allowed
us to identify in great detail the places where
SF is taking place and the role played by the different phases of the interstellar medium
in setting the star formation rate (SFR). As a result the field
of SF has advanced to a new era in which new, sophisticated theoretical models and simulations have been
developed to understand the triggering mechanisms of SF.

Improved observations include high quality, spatially resolved observations of
atomic hydrogen (HI) \citep[e.g.][]{Walter08} and carbon monoxide ($^{12}\rm CO$; e.g. \citealt{Helfer03};
\citealt{Leroy09}), and of ultraviolet (UV) and infrared (IR) SFR tracers in samples
of nearby galaxies.
These data have allowed the accurate estimation of the molecular and atomic hydrogen 
contents of galaxies over a wide range of morphologies and gas fractions.
In addition, more reliable estimates of the unobscured SF
in the UV \citep[e.g.][]{Gildepaz07} and of dust-obscured SF
in the IR \citep[e.g.][]{Calzetti07} have allowed better
determinations of the SFR, both globally and within
individual SF regions. The combination of measurements of spatially resolved 
gas contents and SFR in galaxies led to a better characterisation
of the SF law (i.e. the relation between the surface density of SFR and
gas). There is now support
for a SF law in which the SFR per unit area, $\Sigma_{\rm SFR}$, and the molecular gas
surface density, $\Sigma_{\rm mol}$, correlate linearly, $\Sigma_{\rm
SFR}\propto\Sigma_{\rm mol}$ (\citealt{Wong02};
\citealt{Bigiel08}). This relation is much
stronger than that using either the surface densities of total cold gas or HI.

Towards the outskirts of galaxies, it has been observed that $\Sigma_{\rm SFR}$ starts correlating
with the atomic gas surface density, $\Sigma_{\rm atom}$,
but this correlation appears to be a result of the underlying correlation between
 $\Sigma_{\rm mol}$ and $\Sigma_{\rm atom}$ (\citealt{Bigiel10}; \citealt{Schruba11}).
Most of the studies probing the $\Sigma_{\rm 
SFR}-\Sigma_{\rm mol}$ correlation rely on $^{12}\rm CO$ observations (hereafter CO). The reliability
of this proxy to trace the bulk of the molecular content of galaxies is compromised in low metallicity
gas, typical of dwarf galaxies and the outskirts of spiral galaxies
(e.g. \citealt{Bell06}). However, recently, the linear correlation
$\Sigma_{\rm SFR}\propto\Sigma_{\rm mol}$ has been
confirmed in low metallicity environments through observations of dust column density
and atomic hydrogen by \citep{Bolatto11}, which are independent of the CO.
At high redshifts, there are indications that the same
form, $\Sigma_{\rm SFR}\propto\Sigma_{\rm mol}$ could hold
(\citealt{Bouche07}; \citealt{Genzel10}).
The correlation of
$\Sigma_{\rm SFR}$ with $\Sigma_{\rm mol}$ seems physically reasonable
since stars are observed to form in dense molecular gas clouds (see
\citealt{Solomon05} for a review on the CO-IR luminosity relation and \citealt{Kennicutt12} for a review on the 
SF law of local galaxies).

All of the observational support
for the strong correlation between $\Sigma_{\rm SFR}$ and $\Sigma_{\rm mol}$
 indicates that SF occurs only where the hydrogen in the ISM
has been converted into molecular hydrogen (H$_2$). With this in mind, it is fair to say
that any realistic galaxy formation simulation should incorporate
molecular hydrogen and the subsequent SF taking place from it.
The relevance of the different ISM phases in determining the
SF, suggests that an understanding of star
formation requires an understanding of the
ISM and the formation of neutral warm (atomic) and neutral cold (molecular) gas phases.

\vspace{-0.2cm}\section{Hydrodynamical simulations including multi-phase ISM}

The characterisation of the SF law of local galaxies has pushed the field of
SF to a new stage in which more and more accurate calculations have been carried out,
in which processes such as H$_2$ formation and destruction, HI to H$_2$ transition in non-equilibrium
chemistry, non-equilibrium thermal state, variations in the strength of the radiation field, and in some cases, radiative
transfer, have been included (e.g. Pelupessy et al. 2006, 2009; \citealt{Robertson08}; 
\citealt{Gnedin09}; \citealt{Dobbs11}; \citealt{Shetty12}; 
\citealt{Glover12}; see \citealt{Klessen09} for a status report on numerical simulations).
In this section, we give a brief overview of the work developed to answer three broad questions.

{\it Is the relation between the surface density of H$_2$ and
the SFR causal or is it the result of both quantities correlating to some third,
more relevant quantity?} Theoretical models have explained the observed relation between the H$_2$ surface density and
SFR surface density as resulting from
an underlying correlation between temperature and the chemical state of the ISM \citep{Schaye04}. 
Although molecular hydrogen is not an important
coolant in galaxies today (which have gas metallicities of $Z>0.1 Z_{\odot}$),
 it is an excellent tracer of cold, high-density gas. This happens since both H$_2$ and
cold gas are sensitive to photo-dissociation from UV photons and are therefore found
in places where the gas has self-shielded to prevent photo-dissociation.
Given the fact that low temperatures are required for further gas fragmentation and SF,
one would expect a correlation between SF and the chemical state of the gas to be present
(i.e. the presence of
H$_2$ molecules). This relation has been recently quantified by \citet{Gnedin11}, \citet{Feldmann11} and 
\citet{Glover12} in which both
the molecular hydrogen formation rate and the gas cooling rate correlate with column density of clouds.
 Glover \& Clark show that molecular hydrogen is not a pre-requisite for SF as the 
low temperatures and high-densities 
needed to form stars are reached even in situations were the gas remains atomic. This might be the case 
for the ISM in low-metallicity galaxies, in which dust, the catalyst in the process of H$_2$ formation, is very scarce.  

In low metallicity environments, thermal equilibrium is also expected to be reached
much faster than chemical equilibrium, with the onset of SF 
taking place before chemical equilibrium is reached (\citealt{Wolfire08}).
The predictions from simulations is that eventually, at very low metallicities, the observed $\Sigma_{\rm SFR}-\Sigma_{\rm mol}$
relation should break down. However, this is expected to happen only at metallicities smaller than
$10^{-2}\,Z_{\odot}$ (e.g. \citealt{Glover12}).
Even in such low metallicity galaxies, \citet{Krumholz12} argues that molecular gas is still a good tracer 
of where SF is taking place, given that a correlation between the time averaged
$\Sigma_{\rm SFR}$ and $\Sigma_{\rm mol}$ emerges after reaching chemical equilibrium.
This result is very important for cosmological galaxy formation simulations, in which the
 timescales involved are
much longer than both the thermal and chemical equilibrium timescales in the ISM. Therefore,
processes taking place on small scales in the ISM are part of the overall simplified physical
treatment of sub-grid physics.

{\it What sustains the turbulence in the ISM?} 
Recent hydrodynamical simulations have included energy injection from SNe to study their effect 
 on ISM turbulence and on galaxy structure with the aim of distinguishing it from turbulence driven by 
gravitational instabilities. Contradictory results have been obtained. \citet{Dobbs11} show that feedback 
has an important effect on setting the scale height of the galaxy disk through its effect on the 
velocity dispersion, in a simulation of a large spiral galaxy, 
similar to the Milky-Way, and including the effect of spiral arms. 
Although Dobbs et al. show that the spiral arms contribute to increasing the rate of cloud 
collisions, driving the formation of larger and more massive molecular clouds in the disk, 
gravitational instabilities do not dominate the turbulence in the ISM. \citet{Shetty12} also find 
that SN feedback is the primarily source of turbulence in the dense environments of 
starburst galaxies. 
In addition, \citet{Acreman12} compared simulated maps of HI with the observations of the Canadian Galactic Plane Survey 
and found that simulations match the observations only when SNe feedback is included.
On the other hand, \citet{Bournaud10} show in a simulation of a dwarf galaxy, with properties similar to the 
Large Magellanic Cloud, that feedback is not important in setting the scale height and that this is set by 
gravitational instabilities (which primarily determine the velocity dispersion in their simulation). 
Nonetheless, Bournaud et al. find that feedback is necessary to replenish the large scale turbulence 
which initiates the cascades to small scales and to suppress the formation of very dense, small gas clumps 
(see also \citealt{Hopkins12b} and \citealt{Shetty12}). 
From the latter it is unclear whether the scale height in the simulation of Bournaud et al. 
is in equilibrium or not. There is still a lot of research to be done to get to the root of these 
discrepancies, such as to study the dependence on galactic environment, on the 
implementation of feedback, and study how the results vary with the details of the 
codes (grid vs. smooth particle hydrodynamics).

{\it Cosmological applications of multi-phase ISM simulations.}
Recently, a number of papers have shown that galaxies with realistic properties can be obtained 
in cosmological simulations with a multi-phase treatment of the ISM and realistic gas density thresholds 
for the onset of SF (e.g. \citealt{Murante10}; \citealt{Guedes11}). 
The general agreement in hydrodynamical simulations is that it is necessary to resolve down to scales 
comfortably below the Jeans length in order to resolve star-forming regions \citep{Schaye04}.

We started this review 
pointing to the importance of cosmology in developing a complete galaxy formation theory. 
The problem can also be tackled in the opposite direction: cosmology should care about the way the ISM and SF 
are modelled. Two examples of this is (i) the number density of absorbers probed in lines-of-sight to backgrounds quasars 
and (ii) the metallicity-dependent SF law developed by \citet{Krumholz09}.
Regarding (i), \citet{Altay10} show that H$_2$ self-shielding is needed in cosmological, radiative transfer calculations 
to predict the sharp break observed in the number density of absorbers of large column 
densities, $N_{\rm H}>7\times 10^{21}\, {\rm cm}^{-2}$, typical of condensed gas in the ISM of galaxies. 
Regarding (ii), for a long time feedback from massive stars 
has been suggested to be a primary way to quench SF in dwarf galaxies (\citealt{Benson03}). However, recently 
the possibility of SF being reduced in dwarf galaxies simply by the fact that major coolants in the ISM are scarce 
have started to be explored. \citet{Kuhlen12} show that the stellar mass in dwarf galaxies is greatly  
reduced if a metallicity-dependent SF law is included \citep{Krumholz09}. This 
effect is manly seen in galaxies with $Z<0.1\, Z_{\odot}$. A lot of work still needs to be done to explore 
the effect the lower SF efficiency driven by lower metallicities 
can have on e.g. the well known problem of the overabundance of low-mass galaxies predicted by simulations and models 
(e.g. \citealt{Crain09}; \citealt{Bower12}). 

\vspace{-0.2cm}\section{Semi-analytic models including a two-phase ISM}

Until recently, the ISM of galaxies in semi-analytic models was treated as a single star-forming phase
(e.g. \citealt{Cole00}; \citealt{Springel01}). The first attempts to predict the separate HI and H$_2$ contents
of galaxies in semi-analytic models postprocessed the output of
single phase ISM treatments
to add this information {\it a posteriori} (e.g. \citealt{Obreschkow09}; \citealt{Power10}). It
was only very recently that a proper fully self-consistent treatment of the ISM and SF in galaxies
throughout the cosmological calculation was made (\citealt{Cook10}; \citealt{Fu10}; \citealt{Lagos10}; 
see \citealt{Robertson08} and \citealt{Dutton09} for examples of non-cosmological models).
 These work have shown that a consistent treatment in the ISM and SF in galaxies is necessary to
make progress in understanding the gas contents of galaxies, for example in the relation between the H$_2$ and HI contents 
with other galaxy properties (e.g. \citealt{Lagos12}; \citealt{Kauffmann12}). 

Semi-analytic models of galaxy formation have included formalisms to model the ISM and SF which assume 
hydrostatic and chemical equilibrium. 
The models of Lagos et al. and Fu et al. implemented two ways to estimate 
the partition between H$_2$ and HI in the ISM of galaxies: (i) the empirical relation of \citet{Blitz06}, which relates the molecular-to-atomic surface density ratio
to the hydrostatic pressure within the disk, estimating the SFR from the molecular
gas surface density using the well measured molecular depletion timescale \citep{Bigiel08} (also used by \citealt{Cook10}), 
and (ii) the theoretical law of \citet{Krumholz09}, which models
SF as taking place in turbulent, marginally stable clouds,
estimating the molecular abundance from the balance between the dissociating radiation flux and the formation of molecules
on the surface of dust grains. 
Both models show that the empirical relation of Blitz \& Rosolowsky allows a good fitting to the 
HI mass function at $z=0$ (\citealt{Zwaan05} and \citealt{Martin10}).
\citet{Lagos11} used an improved resolution Monte Carlo simulation 
to show that the agreement between the model prediction and the observations 
extends down to HI masses of $10^6\, M_{\odot}\, h^{-2}$. This same model predicts a
 clustering of HI selected galaxies in good agreement with observations \citep{Kim12}. 
The inclusion of the Krumholz et al. theoretical law in the Fu et al. and Lagos et al. models leads 
to large overpredictions in 
the number density of intermediate HI mass galaxies. 
This is partially due to a cloud clumping factor (i.e. the ratio between the surface density of clouds 
and the diffuse medium) introduced by Krumholz et al., which is an unknown in galaxy formation models and most simulations.

\begin{figure}[h!t]
\centering
\includegraphics[width=0.39\textwidth]{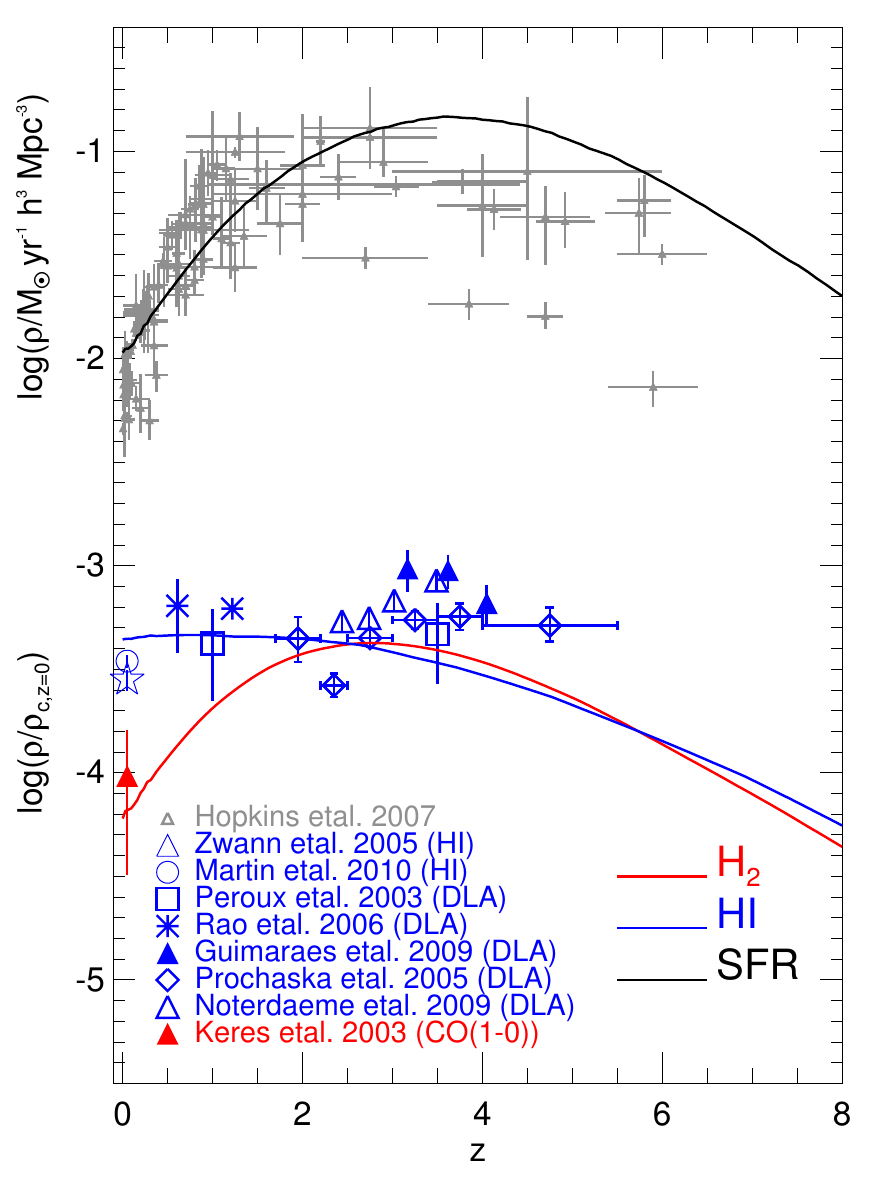}
\includegraphics[width=0.39\textwidth]{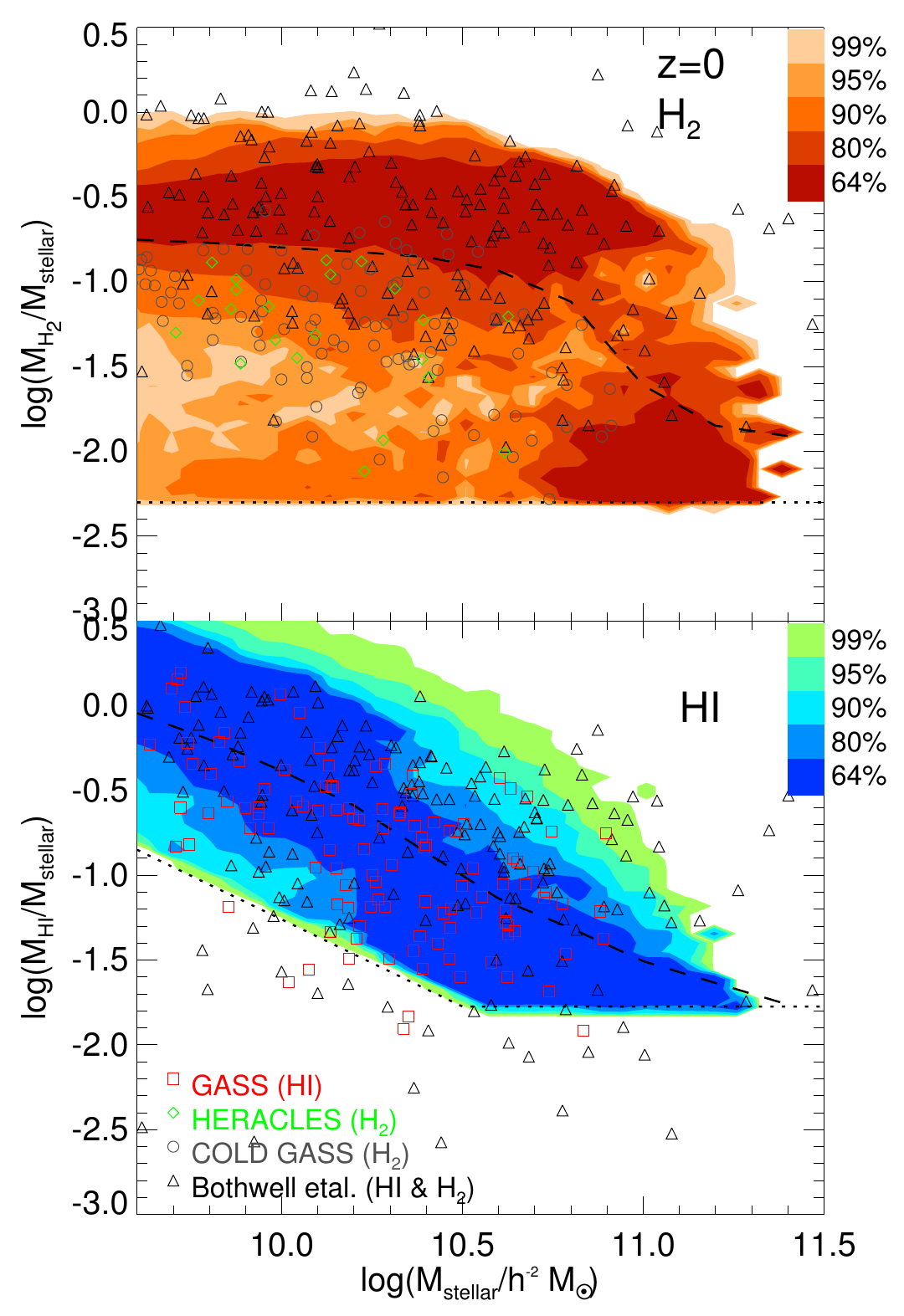} 
\caption{{\it Left panel:} Global density of the SFR in units of $M_{\odot}\, {\rm yr}^{-1}\, h^3\,{\rm Mpc}^{-3}$ (black line), and of the atomic hydrogen (blue line)
and molecular hydrogen (red line) in units of the critical density at $z=0$,
as a function of redshift for the \citet{Lagos11} model. 
The grey crosses correspond to the compilation of observational estimates of the SFR global density of \citet{Hopkins06}.
The HI mass density from \citet{Zwaan05}
and \citet{Martin10} from the HI MF, and \citet{Peroux03},
\citet{Prochaska05}, \citet{Rao06}, \citet{Guimaraes09} and
\citet{Noterdaeme09} from DLAs are also plotted using blue symbols, as labelled. Also shown
is the local Universe estimate of the H$_2$ mass density from \citet{Keres03}
using the $\rm CO(1-0)$ luminosity function as red symbol. {\it Right panel:} the ratios 
H$_2$-to-stellar mass (top panel) and HI-to-stellar mass (bottom panel) 
as a function of stellar mass for the \citet{Lagos11} model at $z=0$.
The dotted line show the approximate sensitivity limits
below which CO(1-0) or HI are not detected in the different surveys.
Contours
show the regions within which
different volume-weighted percentages of the galaxies lie for a
given stellar mass and above the sensitivity limit,
with the scale
shown by the key. {For reference, the dashed line shows the median of the model
distributions.} Observational data
from the HERACLES survey \citep{Leroy09}, the GASS catalogue \citep{Catinella10}, 
the COLD GASS
survey \citep{Saintonge11} and the literature
compilation of \citet{Bothwell09} are shown as symbols.}
\label{fig:DensEvo}
\end{figure}

Two key results obtained from semi-analytic models when a realistic ISM modelling is included are shown in 
Fig.~\ref{fig:DensEvo}. {\it The steep decline of the SFR density with decreasing redshift observed in the Universe 
is closely connected to the steep decline of the molecular mass density}, 
while the atomic hydrogen density is predicted to evolve very weakly 
with redshift (\citealt{Lagos11}). Lagos et al. explain this relation as arising from a combination of 
decreasing gas fractions and increasing galaxy sizes with decreasing redshift. Both act reducing the gas surface density 
and the hydrostatic pressure of the disk. Thus, the SFR density evolution can be linked with the evolution of the surface density of 
gas in the galaxies dominating the SFR in the Universe at each time. 

{\it The increasing hydrostatic pressure in galaxy disks and the increase in gas fractions 
with increasing redshift also lead to an increase in the molecular 
to dynamical mass ratios}. This increase has been inferred in observations of normal 
galaxies in the Universe up to $z\sim 2.5$ (see \citealt{Geach11}). 
\citet{Lagos11} show that the shape of the relation between the molecular gas fraction (i.e. the 
molecular to molecular plus stellar mass ratio) and redshift depends   
on the environment of galaxies, in a way that galaxies residing 
in low mass halos have larger molecular gas fractions than those residing in more massive halos, on average. 
The simulations of \citet{Narayanan12b} and \citet{Dave11} agree qualitatively with this prediction. 
A main problem arising when testing model predictions against observations of molecular gas 
is that observational samples are inhomogeneous.
In order to constraint better the predictions of the 
simulations it is necessary to estimate properties of the environments of the observed galaxies at various redshifts, stressing the 
need for homogeneous, volume-limited samples of galaxies. 

\citet{Saintonge11} presented the first local Universe CO$(1-0)$ volume-limited sample and in combination with the 
HI observations of \citet{Catinella10}, it was possible to study the relations between 
the atomic and molecular gas contents with other galaxy properties. 
This takes us to the second key 
result obtained in semi-analytic models, {\it the molecular-to-atomic gas ratio
correlates with stellar mass, and
that there is an anti-correlation between the HI-to-stellar mass ratio and stellar mass.} 
Fig.~\ref{fig:DensEvo} shows the scaling relations of the gas content with stellar mass. 
The predicted relation between $M_{\rm H_2}$ and the stellar mass, $M_{\star}$, is close to linear 
 for galaxies that lie on the active star-forming sequence in the $M_{\star}-\rm SFR$ 
plane (\citealt{Brinchmann04}). 
{\it What drives the close to constant $\rm SFR/$$M_{\star}$ and
$M_{\rm H_2}/M_{\star}$ ratios for galaxies on the active star-forming sequence is the balance
between accretion and outflows, mainly regulated by the timescale for gas to be reincorporated
into the host halo after ejection by SNe} (\citealt{Lagos10}). 
It is precisely the evolution of this balance that causes the decline in the global molecular gas density with decreasing 
redshift shown in the left panel of Fig.~\ref{fig:DensEvo}.
\citet{Kauffmann12} show that an important 
constraint on models is imposed by the non-detections obtained in the volume-limited surveys since they represent 
quenched galaxies, i.e. galaxies with small gas reservoirs. Kauffmann et al. show that models do not easily reproduce the 
observed trends between the fraction of galaxies detected in CO$(1-0)$ emission and galaxy concentration, stellar mass and  
surface density. A possible explanation for this 
is that semi-analytic models do not take into account gas flows in the ISM and dynamical drivers, such as bars, which 
would require better modellings of the morphological transformation of galaxies.

\begin{figure}[h!t]
\centering
\includegraphics[width=0.775\textwidth]{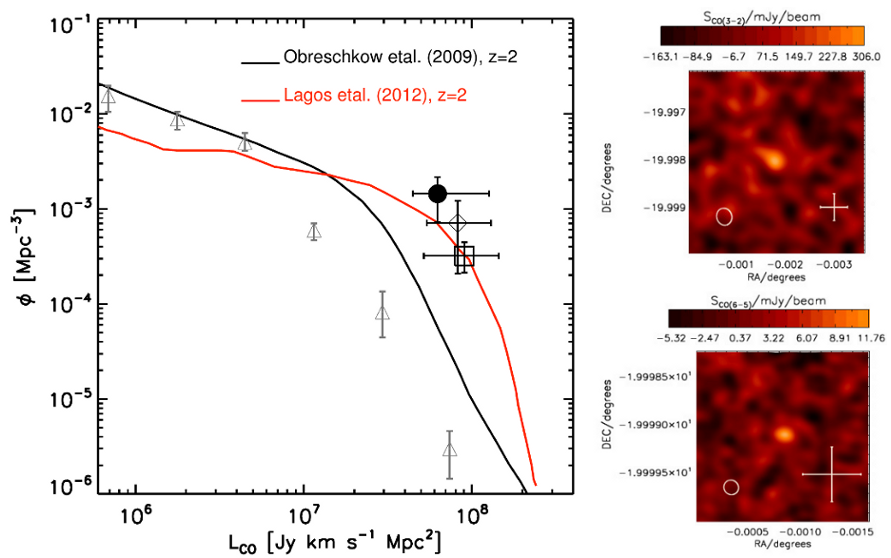}
\caption{{\it Left panel:} The $\rm CO(1-0)$ luminosity function inferred in observations by \citet{Keres03} at $z=0$ (open triangles) and 
\citet{Aravena12} (filled circle and open diamond) and \citet{Daddi10} (open square) at $z=2$. 
The predictions from the two semi-analytic models, Obreschkow et al. (2000; black line) and 
Lagos et al. (2012; red line), 
which include a physical calculation of the CO emission in the ISM 
of galaxies are also shown (Fig. credit for Manuel Aravena). {\it Right panel:} simulated 
observations of the CO$(3-2)$ and CO$(6-5)$
flux maps in declination vs. right ascension of a typical star-forming galaxy at $z=2$ predicted by
the \citet{Lagos12} model. Flux is in units of $\rm mJy/beam$ and the flux scale is shown at the top of each panel.
Maps correspond to hypothetical observations of the full ALMA configuration
($50$ antennae) band~3 after $20$~mins (top panel) and band~6 after $5$~hours of integration (bottom panel).
Ellipses at the bottom-left corner
indicate the beam size and shape and the cross shows
1$\times$1~arcsec$^2$.}
\label{CO10z2LF}
\end{figure}

Observations of the molecular gas at high-redshift are very scarce and limited to individual examples. 
The situation with observations is changing very rapidly with the development of more powerful millimeter and radio telescopes. 
An example of this is the recent observational campaign with the JVLA instrument presented by \citet{Aravena12}, 
which follow up the COSMOS field in CO$(1-0)$ 
emission, which in addition to the previous estimate of
\citet{Daddi10}, add important constraints to the CO$(1-0)$ luminosity function at high-redshift (see symbols in Fig.~\ref{CO10z2LF}). 
These campaign suggest a strong evolution in the number density of bright CO$(1-0)$ galaxies from $z\sim 0$ to $z\sim 2$. 
The few constraints on the high-redshifts CO$(1-0)$ luminosity function already help to ruled out some of the 
available models. Physical models for the CO emission in galaxies have started to be explored in cosmological 
simulations with encouraging results, such as the one shown in Fig.~\ref{CO10z2LF} and the CO-IR luminosity relation 
(\citealt{Lagos12}; \citealt{Narayanan12}). 

\vspace{-0.4cm}\section{Conclusions}

We have discussed two broad theoretical approaches to study galaxy formation and evolution: 
hydrodynamical simulations, or ``parallel approach'' and semi-analytic models, or 
``serial approach''. The main advantage of the former is the study of complex phenomena in detail 
and the development of physical models describing for example SF and the phases of the ISM. 
The main advantage of the latter is the extensive testing of the parameter space and 
the broad comparison with observations through large statistics. 

Regarding hydrodynamical simulations, we discuss recent findings 
that explain the observed relation between $\Sigma_{\rm SFR}$ and 
$\Sigma_{\rm mol}$ as arising from an underlying 
relation of both quantities with the surface density of gas and the effect of SNe feedback 
on sustaining the turbulence in the ISM of galaxies. We also give two examples of the importance of the ISM modelling 
 in key cosmological questions. 
Regarding semi-analytic models, we show that the inclusion of hydrostatic, chemical equilibrium descriptions 
for the ISM and SF help the models explain the observed HI and H$_2$ mass function and the observed 
scaling relation of the atomic and molecular gas with other galaxy properties, and the 
observed trend of increasing molecular gas fraction with redshift. 
We have also discussed open problems and 
lines of investigation both techniques are now starting to explore.  

So far, semi-analytic models and simulations of galaxy formation have a plethora of predictions 
of the relation between the H$_2$ and HI gas contents with galaxy properties, and even the CO excitation levels of 
galaxies at different cosmic epochs. The dramatic improvement in the quality and quantity of data
expected over the next decade with the
next generation of radio and sub-millimeter telescopes such as the Australian SKA
Pathfinder, the Karoo Array Telescope 
and the Square Kilometre Array which aim to detect $21$ cm emission from HI,
and the Atacama Large Millimeter Array, the Large Millimeter Telescope 
and the Cornell Caltech Atacama Telescope, which are designed to detect emission from molecules and dust, 
will help constrain and improve the models discussed in this review. 

\nocite{Pelupessy06}
\nocite{Pelupessy09}
\nocite{Obreschkow09d}
\nocite{Lagos12}
\vspace{-0.3cm}\section*{Acknowledgements}

CL thanks Inti Pelupessy, Giuseppe Murante, Claire Dobbs and Peter Creasey for the discussions and help in understanding better 
hydrodynamical simulations. 
CL is grateful to the organizers of the symposium for the invitation and the exciting discussions on the ISM 
of galaxies and the challenges the field currently faces.

\bibliographystyle{mn2e_trunc3}
\bibliography{Simulations_Modelling_ISM}

\end{document}